\documentclass[11pt]{amsart}

\usepackage{graphicx,hyperref,verbatim}
\usepackage[top=26mm, bottom=26mm, left=32mm, right=32mm]{geometry}
\usepackage[dvips]{color}
\usepackage{bm}

\usepackage[T1]{fontenc}
\usepackage{lmodern}

\newcommand{\vct}[1]{\bm{#1}}
\newcommand{\mtx}[1]{\mathsf{#1}}

\numberwithin{equation}{section}
\numberwithin{figure}{section}
\theoremstyle{definition}

\numberwithin{remark}{section}

\numberwithin{definition}{section}

\newcommand{\lsp}{\vspace{3mm}}

\begin{document}

\begin{center}
\textbf{\large Algorithm \texttt{librla}: A library of randomized linear algebra routines}

\lsp

{\small Adrianna Gillman\\
Department of Applied Mathematics, University of Colorado Boulder \\
\lsp

Zydrunas Gimbutas\\
National Institute of Standards and Technology (NIST)}

\lsp

\textbf{Abstract:} The library \texttt{librla} is a randomized linear algebra library that is specifically designed for the intermediate-sized matrices (of dimension up to roughly 10,000) that arise in applications such as reduced order modeling, fast direct solvers, least squares solves and, in some settings, data compression.  \texttt{librla} is the first software package that is both stable and efficient in several high-level languages: MATLAB, Python and Julia.  It also provides increased functionality over existing software.  Specifically, it allows the user to choose to create a factorization based on a fixed rank or a desired tolerance.  The factorization options include QR, SVD and the interpolative decomposition.  Additionally, the factorization can be generated either with access to the matrix or access to a matrix-vector multiplication routine.  Numerical results compare the Python implementation with the available PyTorch and SciPy randomized factorizations.  Performance of \texttt{librla} in the three languages is comparable.

\end{center}

\section{Overview}

Randomized linear algebra for creating low-rank factorizations has become a vital tool in the development of algorithms for neural networks, fast direct solvers and reduced order modeling.  
Additionally, randomized linear algebra provides useful routines for solving total least squares problems and rank-deficient least squares problems, doing matrix approximation, and skeletonizing (i.e., subset selection) a matrix \cite{Chan:1992}.

While there is a large amount of research activity in the field of randomized linear algebra, an easy-to-use, efficient and stable factorization library is not available.  For several factorization libraries, the stability issue manifests itself either as a failure in the form of an invalid factorization or as an inability to return anything.  When new techniques implemented in a high-level language depend on low-level language libraries, this dependence has a big impact on the portability and usability of the new work.  For example, this problem has impacted the field of fast direct solvers for boundary integral equations which often depend on a legacy Fortran interpolative decomposition library \cite{Tygert:2008}.

The library \texttt{librla} (\href{https://github.com/agillman20/librla}{https://github.com/agillman20/librla}), written natively in Python, Julia and MATLAB, provides low-rank QR factorizations, SVDs and interpolative decompositions.  The algorithms randomly sample the range of the matrix or operator in a similar manner to \cite{Halko:2011}. A key feature of this package is that it is designed to exploit Level 3 BLAS operators as much as possible.  The use of Level 3 BLAS allows all the linear algebraic operations in the method to be executed efficiently via low-level optimized code.  \texttt{librla} is designed for small- to mid-range-sized matrices (i.e., of dimension up to roughly 10,000, depending on computing resources).  \texttt{librla} is not intended for matrices that are larger or do not fit in memory.

\texttt{librla} includes the following options, all of which can be used for both real and complex matrices: 

\begin{itemize}
\item \texttt{qr\_sketch}: Randomized QR factorization.
\item  \texttt{svd\_sketch}: Randomized Singular Value Decomposition (SVD).
\item \texttt{id\_sketch}: Interpolative Decomposition via randomized sampling.
\end{itemize}

The user has the choice of specifying a desired accuracy (tolerance) or rank.  For each method, the user has the option to use extra samples and to use a specified number of power iterations.  
The package does include the option to create low-rank factorizations of matrices that are applied via matrix-vector multiplication codes. When the matrix-vector multiplication is ``matrix-free,'' the creation of the low-rank factorization is also matrix-free. In order to use this option, \texttt{librla} requires the ability to apply both the matrix and its conjugate transpose via a subroutine.

The algorithm used in \texttt{librla} is built in a similar manner to the randomized methods in \cite{Halko:2011,Liberty:2007}.  There is a large collection of related work. Review papers on the field include \cite{Martinsson:2020,Mahoney:2011,Drineas2017LecturesOR,Woodruff:2014,Voronin:2017}.  Some literature \cite{Mahoney:2009,Sorensen:2016,Drineas:2008} has focused on the design of CUR factorizations, while other manuscripts \cite{Frieze:2004,Drineas:2006_2,Drinea:2006_3} have focused on Monte Carlo approaches to creating low-rank factorizations.  Randomized projection methods \cite{Halko:2011,Liberty:2007,Duersch:2020,MEIER:2024,Gu:1996,ERICHSON:2018,Rokhlin:2008} are common.  Deterministic factorization techniques exist \cite{Duersch:2017_2,Feng:2018}.  Many techniques are specifically designed for distributed memory  and/or to not require many passes through the data \cite{Martinsson:2019,Sarlos:2006,Tropp:2017,Yu:2017,Lindquist:2020,Yang:2015,Bjarkason:2019}.  There is much literature from the randomized linear algebra community for applications including $L_2$ solutions \cite{Drineas:2006,Drineas:2007,Avron:2010,Meng:2014,Pilanci:2016}, reduced order modeling \cite{Sorensen:2016}, linear programming \cite{Drineas:2012,Chowdhury:2020}, statistics \cite{Drineas:2016}, machine learning \cite{Yao:2021},
Newton methods \cite{Roosta-Khorasani:2019,Pilanci:2017}, and differentially private matrices \cite{Balu:2016,Cormode:2018,Choi:2020}.

\section{Mathematical Foundations}

The algorithm that serves as the foundation of \texttt{librla} is found in the subroutine called \texttt{orth\_sketch}. It creates an orthogonal basis of the range of the operator of interest.  The user-specified tolerance portion of the package was heavily influenced by the methods presented in \cite{Halko:2011,Tygert:2008,Liberty:2007}.  The user-specified rank portion of the package was influenced by the \texttt{svd\_lowrank} routine in PyTorch \cite{Paszke:2019}.  Once the orthogonal basis is created, it is possible to create the low-rank QR, SVD or interpolative decomposition via standard techniques.  A pseudocode of \texttt{orth\_sketch} is presented here. The git repository includes pseudocodes for all of the factorizations.


When calling the desired factorization subroutine, the user provides a linear operator $\mtx{A}$ of size $m\times n$ and a desired tolerance  or rank.  The user has the option to use power iteration and to specify the number of iterations.  The power iteration accelerates convergence but involves additional matrix-matrix multiplications. Users are also provided the option to specify a number of oversampling vectors.


Roughly speaking, given a linear operator $\mtx{A}$, a tolerance $tol$ or desired rank $k$, \texttt{orth\_sketch} randomly samples the range of $\mtx{A}$ to create an orthogonal basis for the range. A specified block size $b$, number of power iterations $q$ and number of oversampling vectors $p$ are optional inputs. For simplicity of presentation, in this paragraph, set $p$ and $q$ equal to zero.  The range is sampled by applying $\mtx{A}$ to a matrix $\Omega$ of size $n \times b$ whose entries are sampled uniformly at random from $[-1,1]$.  This choice of matrix entries for $\Omega$ ensures that the vectors are linearly independent; while the optimal choice of sampling distribution is an open question, the choice is not critical. (The error bound in Lemma~4.1 of \cite{Halko:2011} holds for uniform entries with a slightly different constant.)  The orthogonal basis is formed by taking a pivoted QR factorization of $\mtx{A}\Omega$, where $\Omega$ is optionally refined via power iteration with re-orthogonalization at each step.  The magnitudes of the diagonal entries of $\mtx{R}$ ($|{\rm diag}(\mtx{R})|$) are used in the stopping criterion of the tolerance mode.  \texttt{orth\_sketch} returns $\mtx{Q}$ or a submatrix of $\mtx{Q}$ and $|{\rm diag}(\mtx{R})|$. The matrix $\mtx{Q}$ is then used to create the desired factorization.  The number of columns of $\Omega$ and the acceptance test are set differently in the fixed rank and tolerance modes, as described next.

In the fixed rank mode, the sample matrix $\Omega$ is drawn with $b = k + p$ columns, where $k$ is the requested rank and $p$ is the number of oversampling vectors (the optional argument \texttt{extra\_samples} is set to 12 by default).  The approximation error is sensitive to the decay of the trailing singular values.  The small number of extra samples improves the expected error and makes the probability of a large deviation from the true spectrum smaller \cite{Halko:2011,Martinsson:2020,Tropp:2017}. When there is slow decay of the singular values, oversampling combined with power iteration improves the accuracy as it sharpens the spectral decay of the sketched operator.  This is illustrated in Figure \ref{fig:climate_singular}.  The same strategy of augmenting the requested rank with extra samples and power iteration is employed by PyTorch's \texttt{svd\_lowrank} \cite{Paszke:2019}.



In the tolerance mode, the initial sketch is drawn with the default block size (of $\Omega$) set to 42 columns. The user can change this value through an optional argument.  The range is determined to be sufficiently captured when the ratio of the smallest and largest (in magnitude) diagonal entries of $\mtx{R}$ is below the specified tolerance.  If the stopping criterion is not satisfied, the block size is increased by a factor of 4 and a new sketch is drawn.  Because the block sizes grow geometrically, the cumulative computational cost of all rounds is bounded by a small constant multiple of the computational cost of the final round alone.  When the power iteration is used, the computational cost for each increase of the block size increases proportionally with the size but the constant prefactor does not change.  If the tolerance is never satisfied, the loop aborts before the block size reaches 
$\min\{m,n\}$ and the calling routine falls back to a deterministic full pivoted QR factorization. Thus the total worst-case computational cost remains a small constant multiple of the cost of a full pivoted QR factorization.

By choosing the stopping condition to be evaluated via the magnitude of the entries of the matrix $\mtx{R}$ resulting from the QR factorization of $\mtx{A}\Omega$ which creates the orthogonal basis for the range, there is no computational cost associated with evaluating it.  This choice of stopping condition coincides with the standard a posteriori error estimate for the randomized range finder \cite{Woolfe:2008,Halko:2011,Martinsson:2020}. The pivot chosen at step $k+1$ of the QR factorization is the residual column of largest norm, $|r_{k+1,k+1}| = \max_j \|(\mtx{I}-\mtx{Q}_k\mtx{Q}_k^*)\mtx{A}\omega_j\|_2$ where $\omega_j$ is the $j^{\rm th}$ column of $\Omega$ and $\mtx{Q}_k$ denotes the orthogonal matrix at the $k^{\rm th}$ step of the QR factorization.  Thus the sample columns beyond the index where the tolerance is first met act as random test vectors whose residual norms all lie below the threshold, indicating that, with high probability, $\|(\mtx{I}-\mtx{Q}_k\mtx{Q}_k^*)\mtx{A}\|$ is correspondingly small. The estimate is optimistically biased \cite{Epperly:2024}.  Thus, the stopping indicator is heuristic rather than a guaranteed bound.

\begin{figure}[ht]
\fbox{
\begin{minipage}{140mm}
\begin{center}
\textsc{Algorithm}: Pseudocode for the \texttt{orth\_sketch} algorithm
\end{center}
The \texttt{orth\_sketch} algorithm is the foundation for creating the low rank factorizations via randomized sampling of the range.

\rule{\textwidth}{0.5pt}

    \textbf{Input}: a linear operator $\mtx{A}$ of size $m\times n$, 
            \hspace{0.1cm} a desired rank $k$ or tolerance $\texttt{tol}$\\
           \hspace{0.1cm}  number of power iterations $q$ (optional), number of oversampling vectors $p$ (optional) \\
             desired block size $b$ (optional for tolerance mode only, must satisfy $b>p$) \\
             If $b$ is not provided, it is set to $42$.  If $q$ is not provided, it is set to $0$.  If $p$ is not provided, it is set to $12$.\\
    \textbf{Output}: $\mtx{Q}$ (orthonormal basis), $|{\rm diag}(\mtx{R})|$  \\
\begin{tabbing}
\mbox{}\hspace{7mm} \= \mbox{}\hspace{6mm} \= \mbox{}\hspace{6mm} \= \mbox{}\hspace{6mm} \= \mbox{}\hspace{6mm} \= \kill

\> \textbf{if} a rank $k$ approximation is desired  \\
\>\>    $b = k + p$ \\
\>\>    Create the uniformly sampled $[-1,1]$ random matrix  $\mtx{\Omega}$ of size $n \times b$ \\
\>\>   Apply power iteration if desired; i.e. $\mtx{\Omega} =(\mtx{A}^* \mtx{A})^q \mtx{\Omega}$ \\
\>\>      Set  $\mtx{Y} = \mtx{A \Omega}$ \\
\>\>      Let $\mtx{Q}$ and $\mtx{R}$ denote the factors that result from the QR factorization of $\mtx{Y}$\\
\>\>        return  $\mtx{Q}$, $|{\rm diag}(\mtx{R})|$ \\ 
\>\>\>(the orthogonal basis $\mtx{Q}$ will be truncated to rank $k$ outside)\\
\> \textbf{end if}\\
\> \textbf{if} tolerance mode is desired \\
\>\>    Loop until exit: \\
\>\>\>    Create the uniformly sampled $[-1,1]$ random matrix  $\mtx{\Omega}$ of size $n \times b$ \\
  \>\>\>     Apply power iteration if desired; i.e. $\mtx{\Omega} =(\mtx{A}^* \mtx{A})^{q} \mtx{\Omega}$ \\
\>\>\>       Set  $\mtx{Y} = \mtx{A \Omega}$ \\
\>\> \>      Let $\mtx{Q}$ and $\mtx{R}$ denote the factors that result from the QR factorization of $\mtx{Y}$\\
\>\>\>  Truncate $\mtx{R}$ so the smallest dimension is $b-p$ \\
\>\> \>     Define tolerance indicator: \\
\>\> \>  Set $r_{max}$ to the largest in magnitude diagonal entry of $\mtx{R}$\\
\>\> \>  Set $r_{min}$ to the smallest in magnitude diagonal entry of $\mtx{R}$\\
\>\> \> Set $\mathrm{del} = \frac{r_{min}}{r_{max}}$  \\
\>\> \>   Check the indicator:\\
\>\> \>    \textbf{if} $\mathrm{del} \leq \texttt{tol}$: \\
\>\> \> \> If tolerance is met, exit\\
\>\> \> \>            return $\mtx{Q}$,  $|{\rm diag}(\mtx{R})|$  \\
\>\> \> \textbf{end if}\\
\>\> \>     If tolerance is not met, increase block size; i.e.  $b = \min(4 \times b, \min(m, n))$\\
\>\> \>     \textbf{if} $b \geq \min(m, n)$: \\
\> \>\> \> if the block size $b$ is bigger than the smallest dimension of $\mtx{A}$, exit.\\ 
\> \>\> \> Tolerance condition is not achievable.\\
\> \>\> \>            return empty($m, 0$), $[]$  \\
\>\> \> \textbf{end if}\\
\> \textbf{end if}\\
\end{tabbing}
\end{minipage}}
\end{figure}

\section{The Different Factorization Options}

There are three different factorization options.  This section provides details on what the different factorizations accomplish.  The factorization options are:

\begin{itemize}
\item \textbf{ QR factorization via  \texttt{qr\_sketch}:} 

The subroutine returns two matrices: $\mtx{Q}$ and $\mtx{R}$, and a vector $\vct{piv}$.  The rank of the factorization is $k$, where $k\leq \min\{m,n\}$. The $n$ entries of $\vct{piv}$ are the column pivots.  The matrix $\mtx{Q}$ is of size $m \times k$ and the columns form an orthonormal basis for the range of $\mtx{ A}$.  The matrix $\mtx{R}$ is an upper triangular $k\times n$ matrix. The factorization satisfies

$${\mtx{A}}(:,\vct{piv}) \sim \mtx{Q} \ \mtx{R}.$$

\item \textbf{ SVD via \texttt{svd\_sketch:}} The subroutine returns two matrices: $\mtx{ U}$ and $\mtx{V}$, and a vector $\vct{s}$.  The rank of the factorization is $k$, where $k\leq \min\{m,n\}$.  The vector $\vct{ s}$ has $k$ entries that are the singular values of $\mtx{ A}$.  The matrix $\mtx{U}$ is of size $m \times k$ and contains the left singular vectors of $\mtx{ A}$.  The matrix $\mtx{V}$ is of size $n\times k$ and contains the right singular vectors of $\mtx{A}$.  The columns of both $\mtx{U}$ and $\mtx{V}$ are orthonormal.  The factorization satisfies

$$ \mtx{A}  \sim  \mtx{U}\,{\tt diag}(\vct{ s})\,\mtx{V}^{*},$$

where ${\tt diag}(\vct{ s})$ is a diagonal matrix with non-zero entries coming from the vector $\vct{s}$.

\item \textbf{Interpolative factorization via \texttt{id\_sketch}:} The subroutine returns the number of skeleton columns $k$, a vector $\vct{ piv}$ of size $1\times n$ and a matrix $\mtx{T}$ of size $k \times (n-k)$.  The first $k$ entries of $\vct{ piv}$ denote the skeleton columns; the remaining entries stay in natural order.  The matrix $\mtx{T}$ is called the interpolation matrix.  The approximation satisfies the following:

\begin{equation}
\label{eq:id}
\mtx{A}(:, \vct{piv}(k+1:end))\sim\mtx{A}(:, \vct{piv}(1:k))\, \mtx{T}.
\end{equation}

To recover an approximation of the full matrix, the extended interpolation matrix $\mtx{ W}$ of size $k \times n$ is built by:
 $$\mtx{W}(:,\vct{piv}(1:k)) = \mtx{ I}_k,$$
  where $\mtx{ I}_k$  is an identity matrix of size $k$ and 
$$\mtx{W}(:,\vct{piv}(k+1:end)) = \mtx{T}.$$  
The result is that

$$\mtx{ A} \sim \mtx{A}(:, \vct{ piv}(1:k))\, \mtx{ W}.$$
\end{itemize}

The interpolation matrix $\mtx{T}$ is constructed by considering the matrix $\mtx{R}$ that results from the 
QR factorization from \texttt{qr\_sketch} which satisfies $\mtx{A}(:,\vct{piv}) \sim \mtx{Q}\,\mtx{R}$, where 
$\vct{piv}$ is defined above.  The first $k$ entries of $\vct{piv}$ are called the \emph{skeleton} columns.  
The matrix $\mtx{R}$ is partitioned as follows: $\mtx{R} = [\mtx{R}_{11}\ \mtx{R}_{12}]$ where the $k\times k$ upper triangular block $\mtx{R}_{11}$ corresponds to the skeleton columns.
  The interpolation matrix can then be computed in three ways, selected via the \texttt{method} argument.  The default (\texttt{fast}) computes $\mtx{T} = \mtx{R}_{11}^{-1}\mtx{R}_{12}$ by a triangular solve. This is the cheapest option and requires no further access to $\mtx{A}$.  The second option (\texttt{svd}) computes the minimum-norm solution $\mtx{T} = \mtx{R}_{11}^{\dagger}\mtx{R}_{12}$ from a singular value decomposition of $\mtx{R}_{11}$, discarding singular values below the specified tolerance (floored at machine precision in the fixed rank mode).  This keeps the entries of $\mtx{T}$ finite when $\mtx{R}_{11}$ is ill-conditioned, for instance when the requested rank exceeds the numerical rank of $\mtx{A}$; if \texttt{fast} detects a numerically singular $\mtx{R}_{11}$, it falls back to this option automatically.  The third option (\texttt{lstsq}) obtains $\mtx{T}$ by solving \eqref{eq:id} in the least-squares sense, using the original columns of $\mtx{A}$ rather than the sketched triangular factor.  It yields the most accurate interpolation coefficients but at the highest cost.  Additionally, it needs explicit access to the columns of $\mtx{A}$.  If $\mtx{A}$ is given as a matrix-free operator, columns of $\mtx{A}$ are extracted one column at a time via additional applications of the operator.

\section{Numerical Examples}

The numerical results presented in this section illustrate the performance of the \texttt{librla} library. 
To illustrate the speed of the algorithms, Section \ref{sec:comp} compares the performance of the Python codes in \texttt{librla} with Python counterparts in PyTorch and SciPy.  The comparison is performed in Python as these are the maintained open source libraries that the authors are aware of.  Section \ref{sec:example} illustrates the capabilities of \texttt{librla} on problems that show up regularly in the randomized linear algebra literature.

All experiments in this section were run on a MacBook Pro with an M4 Max processor and 64 GB of RAM in the virtual environment created by \texttt{setup\_venv.sh}.

\subsection{Comparison with PyTorch and SciPy}
\label{sec:comp}
Currently, there are two related routines in widely available, maintained Python packages.  They are PyTorch's \texttt{svd\_lowrank} and SciPy's \texttt{interp\_decomp} found in the \texttt{scipy.linalg.interpolative} library.  Codes allowing the user to compare the performance are available in the \texttt{compare} folder of our repository.  To illustrate the performance of \texttt{librla}, a subset of the provided examples is presented here.      The experiments consider the following matrices: Hilbert matrices, a matrix where the spectrum decays $\sim1/k$ for integer $k$ between 1 and the size of the matrix, and a matrix from a Gaussian mixture model \cite{Dong:2025}. Table \ref{tab:py} reports a subset of results obtained by running \texttt{compare\_svd\_torch.py}.  For a sequence of matrices, rank 15 approximate singular value decompositions are computed.  This comparison is only for the fixed rank factorizations as that is the only option available in the PyTorch \texttt{svd\_lowrank} software.  Additionally, the singular value decomposition is the only factorization available in PyTorch.  The results demonstrate that the timings are comparable.  Table \ref{tab:scipy} reports a subset of the results obtained by running \texttt{compare\_id\_scipy.py}.  The interpolative decomposition is the only factorization available in SciPy and it allows the user to input a desired rank or tolerance.  The results in the table include both types of experiments.  Table \ref{tab:scipy} also reports the ratio of the time it takes SciPy to compute the factorization over the time it takes \texttt{librla} under the title \textit{speedup}.   The results illustrate the benefits of using the \texttt{librla} package.  

In addition to the performance in terms of speed, the library \texttt{librla} has an additional feature in that it 
is the only package that provides the user the option of three different factorizations: QR, SVD and interpolative decomposition.

\begin{table}[h]
\centering
\begin{tabular}{|l|l|l|l|l|}\hline
 Problem                 & Size      & PyTorch & librla & Relative Error \\ \hline
\hline
 Hilbert matrix           & $2000 \times 1000$ & 0.002s    & 0.0018s  & 3.679e-08  \\ \hline
 Hilbert matrix     & $4000 \times 2000$ & 0.0045s  & 0.0043s  & 1.604e-07      \\ \hline
 Decaying spectrum matrix & $800 \times 600  $ & 0.0026s  & 0.0023s  & 1.529e-01    \\ \hline
 Gaussian mixture model   & $ 400 \times 400$   & 0.0016s & 0.0014s & 7.096e-01      \\ \hline
 \end{tabular}
\caption{\label{tab:py} The table reports the relative error and the time in seconds for creating rank 15 factorizations of different matrices using PyTorch and \texttt{librla}.}
\end{table}

\begin{table}[h]
\centering
\begin{tabular}{|l|l|l|l|l|l|l|}\hline
 Problem                  & Size     & SC & SciPy  & librla  & Speedup & Relative Error \\ \hline
\hline
 Hilbert matrix     & $2000 \times 1000$ & rank 15          & 0.0355s & 0.0024s & 14.6    & 8.571e-08      \\ \hline
 Hilbert matrix        & $4000 \times 2000$ & rank 15       & 0.1557s & 0.0046s & 33.9  & 1.306e-06      \\ \hline
 Decaying spectrum matrix & $800 \times 600 $  & tol  = 0.01        & 0.2933s & 0.0146s & 20      &0              \\ \hline
 Gaussian mixture model   & $400 \times 400 $  &tol = 0.01         & 0.0699s & 0.0055s & 12.8    & 5.659e-05      \\ \hline
 \end{tabular}
\caption{\label{tab:scipy} The table reports the relative error and the time in seconds for creating low-rank factorizations of different matrices using SciPy and \texttt{librla}. The examples involve fixed rank or tolerance approximations. The abbreviation SC means Stopping Condition. }
\end{table}

\subsection{Examples from the randomized linear algebra literature}
\label{sec:example}
To illustrate the ability of the library to handle problems of interest, the techniques are applied to two problems.  The codes for this are available in the repository named \texttt{librla\_applications} (\href{https://github.com/agillman20/librla_applications}{https://github.com/agillman20/librla\_applications}).  The first problem is a data compression problem taken directly from \cite{Tropp:2019}.  The second problem is an image compression problem from \cite{Duersch:2020}.   Examples similar to these are found throughout the randomized linear algebra literature.

\begin{figure}[ht]
\centering
\includegraphics[width=1.0\textwidth]{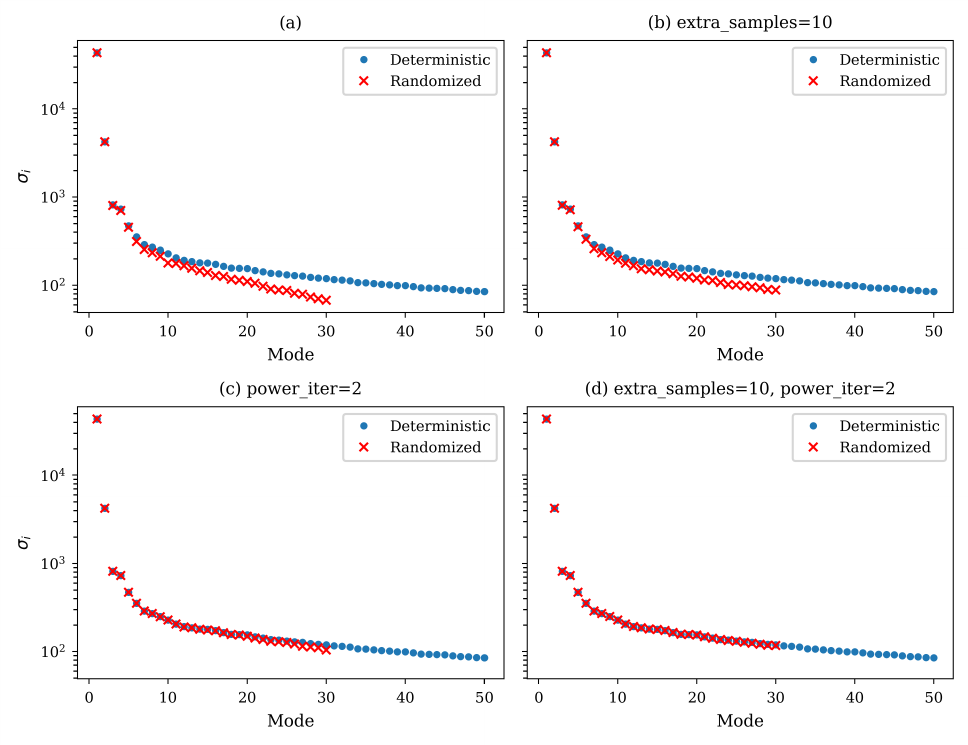}
\caption{Illustration of the exact singular values vs approximate singular values via the (a) randomized SVD, (b) 10 extra samples, (c) 2 power iterations, and (d) 10 extra samples with 2 power iterations.  The example is taken from \cite{Tropp:2019}.  The data is from NOAA Extended Reconstructed Sea Surface Temperature (ERSST), Version 5 \cite{Huang:2017}.  \label{fig:climate_singular}}
\end{figure}

Figure \ref{fig:climate_singular} illustrates the performance of the randomized SVD for NOAA ERSST v5 data \cite{Huang:2017}, following an experiment in \cite{Tropp:2019}. The results were generated by running the \texttt{test\_sst\_modes} code in the \texttt{climate\_analysis} folder with different variations of the option settings.  The specific options used are (a) none, (b) 10 extra samples, (c) 2 power iterations, and (d) 10 extra samples and 2 power iterations.  The extra samples alone do not help the randomized SVD much.  The power iteration is more helpful.  The combination of the two options provides the best approximations of the singular values.  See \cite{Gu:1996} for justification for this improved performance.

\begin{figure}[ht]
\centering
\includegraphics[width=1.0\textwidth]{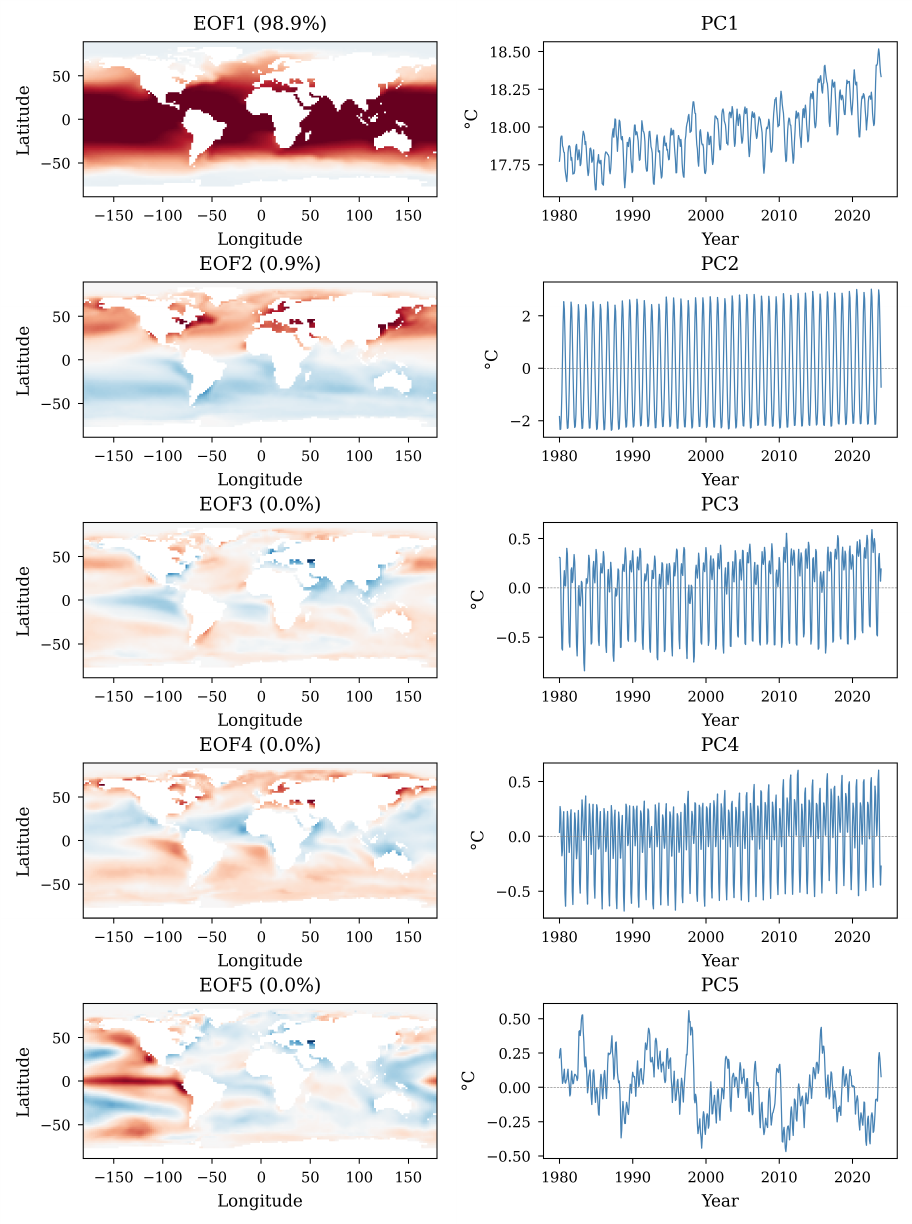}
\caption{Randomized SVD modes of NOAA ERSST v5 sea surface temperature data \cite{Huang:2017}, computed with 10 extra samples and 2 power iterations: spatial empirical orthogonal function (EOF) patterns (left) and principal component (PC) time series (right) for the first five modes.  EOF1 captures the mean spatial pattern, EOF2--4 represent seasonal variations, and EOF5 captures the El Niño–Southern Oscillation (ENSO) signal.  \label{fig:sst_modes}}
\end{figure}

Figure \ref{fig:sst_modes} shows the first five modes of the NOAA ERSST v5 data (empirical orthogonal function (EOF) spatial patterns and principal component (PC) time series), computed with the randomized SVD with 10 extra samples and 2 power iterations.

\begin{figure}[ht]
\centering
\includegraphics[width=1.0\textwidth]{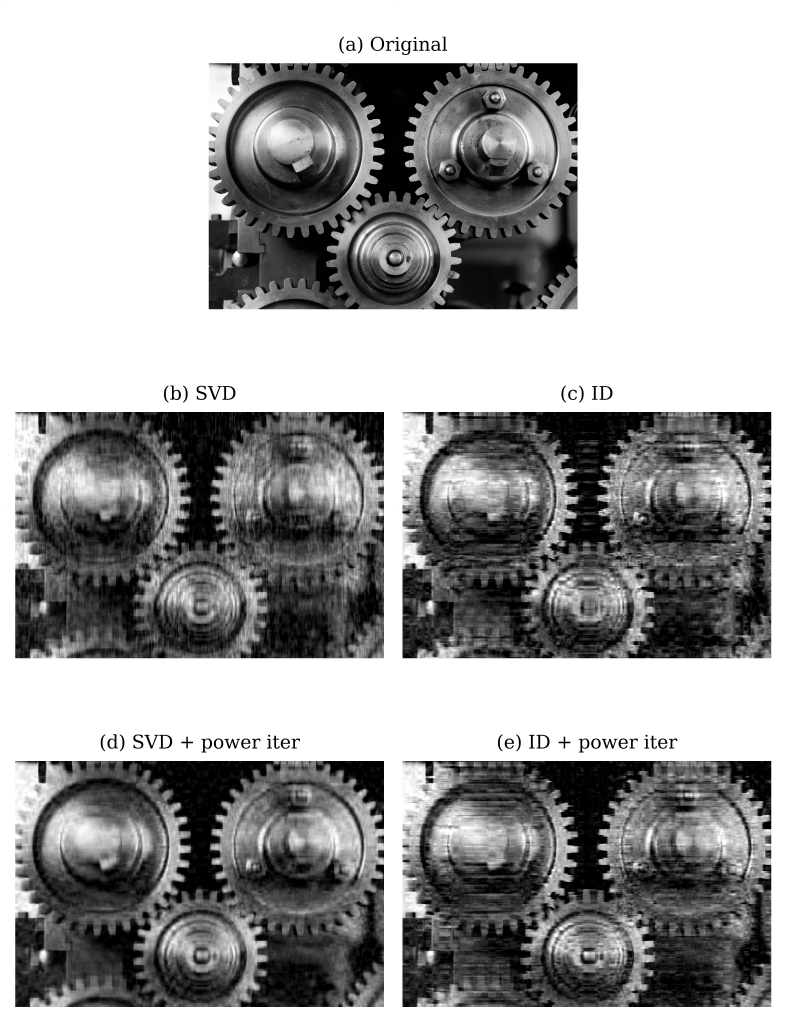}
\caption{Illustration of the use of randomized factorization for image compression: (a) original image, and rank 30 approximations using (b) the randomized SVD, (c) the interpolatory decomposition, (d) the randomized SVD with 15 extra samples and 2 power iterations and (e) the interpolatory decomposition with 15 extra samples and 2 power iterations.  The example is taken from \cite{Duersch:2020}.  The original image is \href{https://www.pexels.com/photo/silver-metal-round-gears-connected-to-each-other-149387/}{``Silver Metal Round Gears''} from Pexels (Creative Commons Zero license).  The image is $4016 \times 6016$ pixels (RGB); it is reshaped into a $4016 \times 18048$ matrix before factorization, and reshaped back for display.  \label{fig:image_orig}}
\end{figure}

The image compression example code \texttt{test\_image\_id} in the \texttt{image\_analysis} folder produces six images.  Figure \ref{fig:image_orig} illustrates five of them: (a) the original image and rank 30 approximations of the image using (b) the randomized SVD, (c) the interpolative decomposition, (d) the randomized SVD with 15 extra samples and 2 power iterations and (e) the interpolative decomposition with 15 extra samples and 2 power iterations.  While the rank of the approximations is the same, the SVD produces an image with less smearing.  The examples with oversampling and power iterations further highlight this result.

\section{Conclusion}
The package \texttt{librla} fills a gap in the software domain by providing efficient and robust randomized factorization techniques in three commonly used high-level languages: MATLAB, Python and Julia.  \texttt{librla} provides more functionality than existing software packages by offering several different factorization options and the option of both fixed rank and tolerance modes.  Additional functionality is offered by allowing users to choose random sample sizes, the number of oversampling vectors and the use of power iteration.  The efficiency of the software comes from the heavy use of Level 3 BLAS computations. A side benefit of the Level 3 BLAS usage is that the codes are very readable.  

Through the GitHub repository \href{https://github.com/agillman20/librla}{https://github.com/agillman20/librla} users can find demo and test codes.  The GitHub repository \href{https://github.com/agillman20/librla_applications}{\texttt{librla\_applications}} demonstrates the
use of \texttt{librla} for compressing data sets that are often seen in the randomized linear algebra literature.

\section{AI usage disclosure}

Generative AI tools were used during the preparation of this work to assist with code development, testing, and editing of the manuscript. The authors reviewed and verified all AI-generated content and take full responsibility for the accuracy and integrity of the final publication. The original algorithm was prototyped and debugged in Python. Claude Code (Anthropic, Claude Sonnet 3.5 and Claude Opus 4.6) was used to assist in porting to Matlab and Julia, while maintaining consistency in API calls and documentation. Unit tests were generated to ensure correctness of the ported code.

\section{Acknowledgements}

The work by A. Gillman was supported by the National Science Foundation (DMS-2110886), and a Knut and Alice Wallenberg Foundation Grant. Part of this work was carried out while A. Gillman was in residence at Institut Mittag-Leffler in Djursholm, Sweden in autumn 2025, supported by the Swedish Research Council under grant no. 2021-06594.

Contributions by staff of NIST, an agency of the U.S. Government, are not subject to copyright within the United States.

Certain commercial software and equipment are identified in this paper to foster understanding. Such identification does not imply recommendation or endorsement by NIST, nor does it imply that the software or equipment identified is necessarily the best available for the purpose.

\bibliography{paper}
\bibliographystyle{amsplain}

\end{document}